\documentclass[jkps,preprint,fleqn,showpacs,showkeys]{revtex4-1}
\usepackage{graphicx}
\usepackage{amssymb}
\usepackage{amsmath}
\usepackage{bm}
\usepackage{mathrsfs}
\begin{document}
\setcounter{page}{0}
\title[]{Mechanical Snell's Law}
\author{KyungTae \surname{Kim}}
\author{June-Haak \surname{Ee}}
\author{Kyounghoon \surname{Kim}}
\author{U-Rae \surname{Kim}}
\author{Jungil \surname{Lee}}
\email{jungil@korea.ac.kr}
\affiliation{Department of Physics, Korea University, Seoul 02841}
\collaboration{\textsl{KPOP}$\mathscr{E}$\footnote{Korea Pragmatist Organization for Physics Education} Collaboration}

\date[]{}

\begin{abstract}
We investigate the motion of a massive particle
constrained to move along a path consisting of two line
segments on a vertical plane
under an arbitrary conservative force. By fixing the
starting and end points of the track and varying
the vertex horizontally, we find the least-time path.
We define the angles of incidence and refraction
similar
to the refraction of a light ray. It is remarkable that 
the ratio of the sines of these angles is identical to
the ratio of the average speeds on the two partial paths
as long
as the horizontal component of the conservative force
vanishes.
\end{abstract}


\maketitle

\section{INTRODUCTION}

Snell's law, which was experimentally discovered 
in about 1621, \cite{Shirley-1951}
states the relationship between
the angles of incidence and refraction
for a light ray crossing a flat interface
of two isotropic media: The ratio of the sines of these
angles is equal to the ratio of the phase velocities
of light in the two media. This relation derives
from Fermat's principle of least time that 
a light ray propagates from a fixed point to another
following the path that minimizes the elapsed 
time. \cite{Feynman-1,Newcomb-1983}
The corresponding derivation is also possible by making use of
calculus of variations \cite{Zatzkis-1964}
or an algebraic approach. \cite{Helfgott-2002}

Invoked by Fermat's principle, calculus of variations
was developed to resolve problems such as the brachistochrone
problem and the shape of a hanging 
chain. \cite{Boyer-2011,Courant-1996}
This progress eventually has led to the establishment of
the Hamilton's principle of least action. \cite{Marion-1970}
The brachistochrone problem is to find the path $P(x,y)$ that minimizes
the elapsed time for a massive particle's motion between two 
fixed points $A$ and $C$ 
on a vertical plane under a uniform gravitational field.
Because the horizontal coordinate
$x=f(y)$ is a function of the vertical coordinate $y$,
the elapsed time $T$ becomes a function of $f$ that is called a functional $T=T[f]$. The function $f(y)$
 can be found by solving the Euler-Lagrange equation
for the functional $T=T[f]$.
The original derivation of Snell's law based on Fermat's
principle does not require calculus of variations
because the determination of the horizontal coordinate $\alpha$
for the intermediate
 point $B$ on the interface that minimizes the elapsed time 
$T(\alpha)$
can be carried
out by taking the derivative with respect to that single parameter 
$\alpha$.


In this paper, we investigate a simplified version of the brachistochrone
problem by imposing a restriction that
a particle slides down
the track consisting of two line segments
$\overline{AB}$ and $\overline{BC}$
under a conservative force without friction.
We keep $A$ and $C$ fixed and vary
the horizontal coordinate $\alpha$ of the vertex $B$.
Due to this constraint, the least-time path
can be found simply by solving the equation 
$dT/d\alpha=0$.
Our analysis
reveals that, as long as
the horizontal component of the conservative force vanishes,
the ratio of sines of the angles of incidence and
refraction, which are defined
similarly to the light-ray refraction, 
is identical to the
ratio of the \textit{average speeds} of the particle on
the two partial paths. We call this relation the \textit{mechanical
Snell's law}. 
To our best knowledge, a realization of Snell's law 
in a purely mechanical system under acceleration is new.

This paper is organized as follows.
In Sec.~\ref{sec:the-generalized-snells-law}, we describe the
mechanical system and derive the mechanical Snell's law.
A formal strategy to compute the corresponding relative index
of refraction under an arbitrary potential is given in
Sec.~\ref{sec:arbitrary-conservative-force}.
 A summary is given
in Sec.~\ref{sec:summary}.

\section{The mechanical Snell's law}\label{sec:the-generalized-snells-law}
In this section, we introduce a mechanical system 
consisting of a particle with mass $m$ constrained to 
slide
along the path consisting of two line segments
$\overline{AB}$ and $\overline{BC}$ on a vertical plane
under a conservative force without friction.
We fix the initial ($A$) and final ($C$) positions and vary the 
vertex $B$ horizontally
to find the point $B=B^*$ at which 
the total elapsed time $T$ from $A$ to $C$ is minimized.
Defining the angles of incidence and refraction
similar to
the light-ray refraction, 
we derive the relation between the two angles.

\subsection{Definitions}
Since the particle's track is on a plane, it is convenient to employ
the two-dimensional Cartesian coordinates:
$A(x_A,y_A)$, $B(x_B=\alpha ,y_B)$, and $C(x_C,y_C)$, where
$x_Q$ and $y_Q$ are the horizontal and vertical coordinates, respectively, of $Q$ for $Q=A,\, B,$ or $C$ with $x_{A}<\alpha<x_{C}$ 
and $y_A>y_B>y_C$.
We denote $v_Q$ by the speed of the particle at point $Q$.
Then the total elapsed time $T(\alpha)$ 
from $A$ to $C$ is a function of a single parameter $\alpha$.
We define $\alpha^*$ by the horizontal coordinate of the vertex  $B^*$.

\begin{figure}
\begin{center}
\includegraphics[width=70mm]{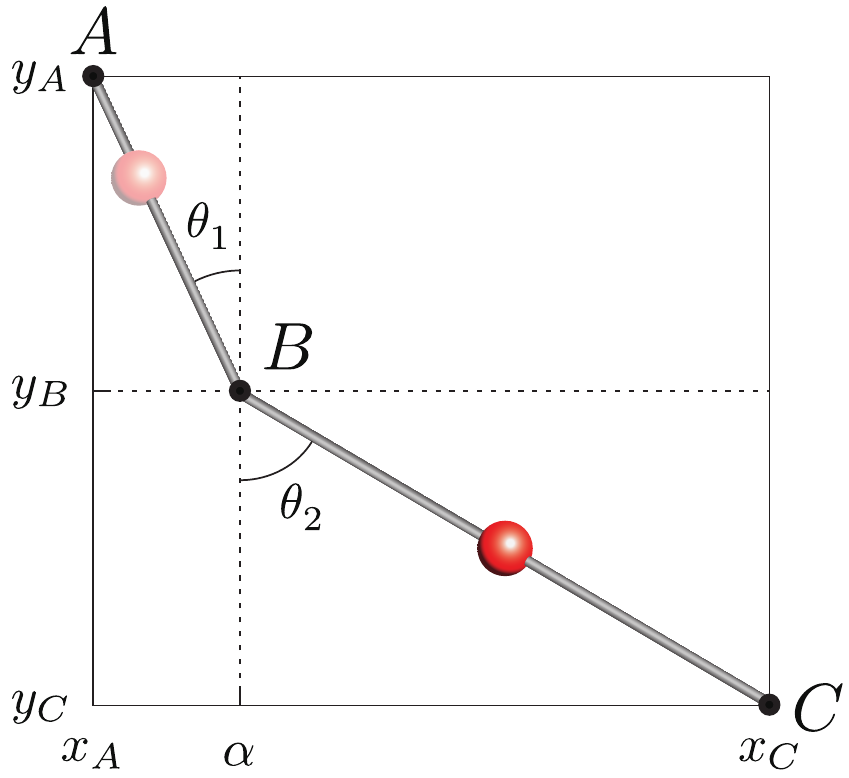}
\caption{\label{fig:model}
A track consisting of two line segments 
$\overline{AB}$ and $\overline{BC}$
on a vertical plane.
A particle of mass $m$ departs from the point
$A$ with the initial speed $v_A$,
slides along the path without friction,
and arrives at the end point $C$ with the terminal speed $v_C$.
$A(x_A,y_A)$ and $C(x_C,y_C)$ are fixed and 
the vertex
$B(x_B=\alpha,y_B)$ varies on the
horizontal line $y=y_B$ with $x_A\le\alpha\le x_C$.
The angle of incidence (refraction) $\theta_1$ ($\theta_2$)
is the angle between $\overline{AB}$ ($\overline{BC}$)
and the vertical.
}
\end{center}
\end{figure}
As shown in Fig.~\ref{fig:model}, 
the angle of incidence (refraction)
$\theta_1$ 
($\theta_2$) is the angle between 
$\overline{AB}$ ($\overline{BC}$) and the vertical.
The angles $\theta_1$ and $\theta_2$ are functions of $\alpha$ as
\begin{equation}
\begin{matrix}
\displaystyle
\theta_1(\alpha)=\arctan\frac{\alpha-x_A}{y_A-y_B},
\\[2ex]
\displaystyle
\theta_2(\alpha)=\arctan\frac{x_C-\alpha}{y_B-y_C}.
\end{matrix}
\end{equation}
Likewise, we use the subscripts $1$ and $2$ to identify 
a physical variable for the paths $\overline{AB}$ 
and $\overline{BC}$, respectively, so that $L_1(\alpha)$,
$T_1(\alpha)$, and $\bar{v}_1(\alpha)$ denote 
the length of $\overline{AB}$,
the elapsed time to pass $\overline{AB}$, 
and the average speed on $\overline{AB}$, respectively.
Thus $L_2(\alpha)$, $\bar{v}_2(\alpha)$, and
$T_2(\alpha)$ represent the corresponding values for $\overline{BC}$.

In summary, 
\begin{eqnarray}
L_1(\alpha)&=&\sqrt{(\alpha-x_A)^2+(y_A-y_B)^2},
\nonumber 
\\
L_2(\alpha)&=&\sqrt{(x_C-\alpha)^2+(y_B-y_C)^2},
\end{eqnarray}
and
\begin{eqnarray}
\cos\theta_1(\alpha)
=\frac{y_A-y_B}{L_1(\alpha)},
\quad
\displaystyle
\sin\theta_1(\alpha)
=\frac{\alpha-x_A}{L_1(\alpha)},
\nonumber \\
\displaystyle
\cos\theta_2(\alpha)
=\frac{y_B-y_C}{L_2(\alpha)},
\quad
\displaystyle
\sin\theta_2(\alpha)
=\frac{x_C-\alpha}{L_2(\alpha)}.
\end{eqnarray}
It is straightforward to express
the average speeds $\bar{v}_1(\alpha)$ and $\bar{v}_2(\alpha)$
as
\begin{eqnarray}
\label{eq:average-speeds-definition}
\displaystyle
\bar{v}_1(\alpha)
=\frac{L_1(\alpha)}{T_1(\alpha)}=
\frac{y_A-y_B}{T_1(\alpha)\cos\theta_1(\alpha)},
\nonumber \\
\displaystyle
\bar{v}_2(\alpha)
=\frac{L_2(\alpha)}{T_2(\alpha)}=
\frac{y_B-y_C}{T_2(\alpha)\cos\theta_2(\alpha)}.
\end{eqnarray}
It is manifest that $\bar{v}_1(\alpha)$ and $\bar{v}_2(\alpha)$
depend on $\alpha$ in general.

\subsection{Least-Time Path}
Let us find the value for $\alpha^*$
which minimizes the total elapsed time,
\begin{eqnarray}
\label{eq:total-elapsed-time}
T(\alpha)
&=&
T_1(\alpha)+T_2(\alpha)
\nonumber \\
&=&
\frac{y_A-y_B}{\bar{v}_1(\alpha)\cos\theta_1(\alpha)}+
\frac{y_B-y_C}{\bar{v}_2(\alpha)\cos\theta_2(\alpha)}.
\end{eqnarray}
We assume that the potential is a well-behaved function
so that 
$T(\alpha)$ is differentiable.
Then the first-order derivative of $T(\alpha)$ 
is given by
\begin{eqnarray}
\label{eq:first-derivative-of-T-intermsof-al}
\frac{d T(\alpha)}{d \alpha}
&=&
\frac{\sin\theta_1(\alpha)}{\bar{v}_1(\alpha)}
-\frac{\sin\theta_2(\alpha)}{\bar{v}_2(\alpha)}
\nonumber \\
&&
-\frac{y_A-y_B}{\bar{v}_1^{2}(\alpha)\cos\theta_1(\alpha)}
\frac{d \bar{v}_1(\alpha)}{d \alpha}
\nonumber \\
&&
-\frac{y_B-y_C}{\bar{v}_2^{2}(\alpha)\cos\theta_2(\alpha)}
\frac{d \bar{v}_2(\alpha)}{d \alpha},
\end{eqnarray}
where we have used
\begin{eqnarray}
\displaystyle
\frac{d}{d \alpha}\left[\frac{1}{\cos\theta_1(\alpha)}\right]
&=&
\frac{\alpha-x_A}{(y_A-y_B)\sqrt{(x_A-\alpha)^2+(y_A-y_B)^2}}
\nonumber \\
&=&
\phantom{-}\frac{\sin\theta_1(\alpha)}{y_A-y_B},
\nonumber \\
\frac{d}{d \alpha}\left[\frac{1}{\cos\theta_2(\alpha)}\right]
&=&
\frac{-(x_C-\alpha)}{(y_B-y_C)\sqrt{(\alpha-x_C)^2+(y_B-y_C)^2}}
\nonumber \\
&=&
-\frac{\sin\theta_2(\alpha)}{y_B-y_C}.
\end{eqnarray}
Because $T(\alpha^*)\le T(\alpha)$ for all $\alpha$, 
we have the constraint
$d T(\alpha)/d\alpha|_{\alpha=\alpha^*}=0$ to find that
\begin{equation}
\label{eq:most-general-angle-speed-relation}
\frac{\sin\theta_1(\alpha^*)}{\sin\theta_2(\alpha^*)}
=
\frac{\bar{v}_1(\alpha^*)}{\bar{v}_2(\alpha^*)}
\,\left[\frac{\displaystyle 1
+\frac{y_B-y_C}{\sin 2\theta_2(\alpha)}
\frac{d \log\bar{v}_2^2(\alpha)}{d \alpha}}
{\displaystyle
1
-\frac{y_A-y_B}{\sin 2\theta_1(\alpha)}
\frac{d \log\bar{v}_1^2(\alpha)}{d \alpha}
}\right]_{\alpha=\alpha^*}.
\end{equation}
If both $\bar{v}_1(\alpha)$ and $\bar{v}_2(\alpha)$ are independent
of $\alpha$, then the relation (\ref{eq:most-general-angle-speed-relation}) collapses into
\begin{equation}
\label{eq:generalized-snell-law-discrete}
n_{12}\equiv
\frac{\bar{v}_1}{\bar{v}_2}
=\frac{\sin\theta_1(\alpha^*)}{\sin\theta_2(\alpha^*)}.
\end{equation}
The relation in 
Eq.~(\ref{eq:generalized-snell-law-discrete}) is
identical to Snell's law
except that the ratio of the phase velocities in two media
is replaced with that of the average speeds 
of the accelerating particle on the two partial tracks.
We call the relation in Eq.~(\ref{eq:generalized-snell-law-discrete}) 
the mechanical Snell's law
and the ratio $n_{12}$ the \textit{relative 
index of mechanical refraction}.

According to Eq.~(\ref{eq:most-general-angle-speed-relation}),
the mechanical Snell's law 
in Eq.~(\ref{eq:generalized-snell-law-discrete})  does not hold 
if $\bar{v}_1(\alpha)$ or $\bar{v}_2(\alpha)$ has an explicit
dependence on $\alpha$.
We shall find in Sec.~\ref{sec:arbitrary-conservative-force} that
the mechanical Snell's law is valid as long as the potential 
depends only on the vertical coordinate so that the 
horizontal force vanishes.

\section{Validity of the mechanical Snell's law}
\label{sec:arbitrary-conservative-force}
In this section, we investigate the requirements of the conservative
force to have the particle's track satisfy the mechanical 
Snell's law in Eq.~(\ref{eq:generalized-snell-law-discrete}).
We first consider an elementary example of the motion 
under a uniform gravitational field.
Next we consider the case involving a general conservative force field
whose potential is independent of the horizontal coordinate.
Both cases are shown to satisfy the mechanical Snell's law.
As the last case, we show that the mechanical Snell's law
in Eq.~(\ref{eq:generalized-snell-law-discrete}) does not hold
but Eq.~(\ref{eq:most-general-angle-speed-relation}) is still valid
if the conservative force has nonvanishing horizontal force.

\subsection{Uniform Gravitational Field}
We consider the case in which a uniform gravitational
field $\bm{g}=-g\hat{\bm{e}}_y$ is applied, where $g$
is the gravitational acceleration and $\hat{\bm{e}}_y$ is
the unit vector along the vertical direction.
In this case, conservation of the total mechanical energy
completely determines the speed of the particle at
$y$ as
\begin{equation}
v(y)=\sqrt{v_A^2+2g(y_A-y)}.
\end{equation}
Hence, $\bar{v}_1(\alpha)$ and $\bar{v}_2(\alpha)$ are 
independent of $\alpha$:
\begin{eqnarray}
\label{eq:constant-gravity-average-speed-12}
\bar{v}_1&=&
\tfrac{1}{2}(v_A+v_B)
\nonumber \\
&=&
\tfrac{1}{2}\big[v_A+\sqrt{v_A^2+2g(y_A-y_B)}\,\big],
\\
\bar{v}_2&=&
\tfrac{1}{2}(v_B+v_C)
\nonumber \\
&=&
\tfrac{1}{2}\big[\sqrt{v_A^2+2g(y_A-y_B)}+\sqrt{v_A^2+2g(y_A-y_C)}
\,\big].
\nonumber \\
\end{eqnarray}
Therefore, the mechanical Snell's law 
in Eq.~(\ref{eq:generalized-snell-law-discrete}) holds.

\subsection{Arbitrary Conservative Force along the
Vertical Direction}
If the conservative force field
is parallel to the vertical direction, then we can define
the potential energy by $V(y)$ which is
independent of the horizontal coordinate $x$ and
the corresponding force is given by
\begin{equation}
\bm{F}=-\bm{\nabla}V(y)=
\displaystyle-\hat{\bm{e}}_y \frac{dV(y)}{dy},
\end{equation}
where $F_y=-dV(y)/dy<0$ and we have
\begin{equation}
v(y)=\sqrt{v_A^2+\tfrac{2}{m}[V(y_A)-V(y)]}.
\end{equation}
By making use of the first 
equality of each line in Eq.~(\ref{eq:average-speeds-definition}),
we can show that
$\bar{v}_1(\alpha)$ and $\bar{v}_2(\alpha)$
are independent of $\alpha$:
\begin{eqnarray}
\label{eq:average-velocity-under-y-potential}
\bar{v}_1
&=&
\displaystyle
\frac{\int_0^{T_1(\alpha)}dt \sqrt{v_A^2+\frac{2}{m}[V(y_A)-V(y)]}}
{\int_0^{T_1(\alpha)}dt }
\nonumber \\
&=&
\displaystyle
\frac{y_A-y_B}
{\int_{y_B}^{y_A}dy\left\{v_A^2+\frac{2}{m}[V(y_A)-V(y)]\right\}^{-1/2}},
\nonumber \\
\bar{v}_2
&=&
\displaystyle
\frac{\int_{T_1(\alpha)}^{T(\alpha)}dt
 \sqrt{v_A^2+\frac{2}{m}[V(y_A)-V(y)]}}
{\int_{T_1(\alpha)}^{T(\alpha)} dt}
\nonumber \\
&=&
\displaystyle
\frac{y_B-y_C}
{\int_{y_C}^{y_B}dy\left\{ v_A^2+\frac{2}{m}[V(y_A)-V(y)]\right\}^{-1/2}},
\end{eqnarray}
where we have changed the integration variable from $t$ to $y$ as
\begin{eqnarray}
\label{eq:dt-dy}
dt
&=&
-\frac{dy\sqrt{1+(dx/dy)^2}}{\sqrt{v_A^2+\frac{2}{m}[V(y_A)-V(y)]}}
\nonumber \\
&=&
-\frac{dy}{\cos\theta_i\sqrt{v_A^2+\frac{2}{m}[V(y_A)-V(y)]}},
\end{eqnarray}
and used the identity $1+(dx/dy)^2=1/\cos^2\theta_i$ with
$i=1$ for $\overline{AB}$ and $2$ for $\overline{BC}$.
The elapsed time $t(y)$ for the particle to reach
the vertical coordinate $y$ can be expressed as
\begin{equation}
\label{eq:t-y-12}
t(y)
=
\begin{cases}
\displaystyle
\frac{1}{\cos\theta_1}
\int_{y}^{y_A}\frac{dy}{\sqrt{v_A^2+\frac{2}{m}[V(y_A)-V(y)]}},&y_A\ge y\ge y_B,
\\[4ex]
\displaystyle
t(y_B)+
\frac{1}{\cos\theta_2}
\int^{y_B}_{y}\frac{dy}{\sqrt{v_A^2+\frac{2}{m}[V(y_A)-V(y)]}},& y_B\ge y \ge y_C,
\end{cases}
\end{equation}
where
 $t(y_A)=0$, $t(y_B)=T_1(\alpha)$, and $t(y_C)=
T(\alpha)=T_1(\alpha)+T_2(\alpha)$.
Because the average speeds in Eq.~(\ref{eq:average-velocity-under-y-potential})
are independent of $\alpha$,
the mechanical Snell's law 
in Eq.~(\ref{eq:generalized-snell-law-discrete}) holds.

\subsection{Arbitrary Conservative Force}
If we consider an arbitrary conservative force field, we can define
the potential energy by $V(x, y)$ 
and the corresponding force is given by
\begin{equation}
\bm{F}=-\bm{\nabla}V(x,y)=
\displaystyle-\hat{\bm{e}}_x \frac{\partial V(x,y)}{\partial x}-\hat{\bm{e}}_y \frac{\partial V(x,y)}{\partial y}.
\end{equation}
Here, we have assumed that $F_x=-\partial V/\partial x<0$ and 
$F_y=-\partial V/\partial y<0$. Thus the speed of the particle at a
point $Q(x,y)$ is expressed as
\begin{equation}
v(x,y)=\sqrt{v_A^2+\tfrac{2}{m}[V(x_A,y_A)-V(x,y)]}.
\end{equation}
Hence, both $\bar{v}_1(\alpha)$ and $\bar{v}_2(\alpha)$
depend on $\alpha$:
\begin{equation}
\label{eq:potential-xy-average-velocity}
\begin{array}{ccc}
\bar{v}_1
&=&
\frac{\displaystyle \phantom{\big(}\alpha-x_A}
{\displaystyle
\int_{\alpha}^{x_A^{\phantom{A}}}dx
\left\{v_A^2+\tfrac{2}{m}[V(x_A,y_A)-V(x,
y_1)]\right\}^{-1/2}}
,
\\
\bar{v}_2
&=&
\frac{\displaystyle \phantom{\big(}x_C-\alpha}
{\displaystyle
\int_{x_C}^{\alpha^{\phantom{A}}}
dx
\left\{v_A^2+\tfrac{2}{m}[V(x_A,y_A)-V(x,
y_2)]\right\}^{-1/2}},
\end{array}
\end{equation}
where $y_1$ and $y_2$ are functions of $x$:
\begin{equation}
\begin{array}{ccc}
y_1(x)&=&y_B+(y_A-y_B)\,\displaystyle\frac{\alpha-x}{\alpha-x_A},
\\[2ex]
y_2(x)&=&y_C+(y_B-y_C)\,\displaystyle\frac{x_C-x}{x_C-\alpha}.
\end{array}
\end{equation}
In deriving Eq.~(\ref{eq:potential-xy-average-velocity}),
we have followed the same procedure to obtain 
Eq.~(\ref{eq:average-velocity-under-y-potential})
except that
we have changed the integration variable from $t$ to $x$ as
\begin{eqnarray}
dt
&=&
\frac{dx\sqrt{1+(dy/dx)^2}}{\sqrt{v_A^2+\frac{2}{m}[V(x_A,y_A)-V(x,y)]}}
\nonumber \\
&=&
\frac{dx}{\sin\theta_i\sqrt{v_A^2+\frac{2}{m}[V(x_A,y_A)-V(x,y_i)]}},
\end{eqnarray}
where we have used the identity $1+(dy/dx)^2=1/\sin^2\theta_i$ with
$i=1$ for $\overline{AB}$ and $2$ for $\overline{BC}$.

It is manifest that, if the conservative force field has non-vanishing
horizontal component, then the mechanical Snell's law
in Eq.~(\ref{eq:generalized-snell-law-discrete}) does not hold 
but the most general form in
Eq.~(\ref{eq:most-general-angle-speed-relation}) should be applied.
This result is consistent with a previous observation in
light-ray refractions:
If the medium does not exert forces parallel
to the interface on the incident photon, then the tangential
momentum of the refracted photon is conserved to satisfy 
Snell's law. \cite{Mooney-1951}

\section{Summary}\label{sec:summary}
We have investigated the relationship between
the angles of incidence and refraction
 of an accelerating particle
sliding down a path consisting of two line segments on a vertical
plane without friction in the presence of an arbitrary conservative force, where the angles are defined similarly to the case of
the light-ray refraction crossing a flat interface of two isotropic media. Our main result which is given in 
Eq.~(\ref{eq:generalized-snell-law-discrete}) states
that, if we choose the least-time path, then
the ratio of the sines of these angles is equal to the ratio of
the average speeds of the particle on the two partial paths
as long as the horizontal component of the conservative force 
vanishes.
The mechanical Snell's law allows us to compute the 
relative index of mechanical refraction
once the
explicit form of the potential is known.

\begin{acknowledgments}
As members of the Korea Pragmatist Organization for Physics
Education (\textsl{KPOP}$\mathscr{E}$), the authors thank
to the remaining members of \textsl{KPOP}$\mathscr{E}$
for useful discussions.

We thank Soo-hyeon Nam, Q-Han Park, and Chaehyun Yu
for their reading the manuscript and useful comments.
This work is supported in part by the National Research Foundation of Korea (NRF) 
under the BK21$+$ program at Korea University, \textit{Initiative for Creative and Independent Scientists}.
The work of JHE, URK, and JL is also supported by the NRF under 
Contract No.~NRF-2017R1E1A1A01074699 (JHE, URK, JL),
NRF-2018R1A2A3075605 (JHE, URK), NRF-2018R1D1A1B07047812 (URK),
NRF-2019R1A6A3A01096460 (URK), and NRF-2017R1A2B4011946 (JHE).
\end{acknowledgments}

\end{document}